\documentclass[reprint,superscriptaddress,amsmath,amssymb,aps,prl,]{revtex4-1}

\usepackage{soul}
\usepackage{graphicx}
\usepackage{dcolumn}
\usepackage{bm}

\usepackage{textcomp}
\usepackage{listings}
\usepackage{float}

\usepackage{color}
\usepackage[usenames,dvipsnames]{xcolor}
\usepackage{siunitx}
\usepackage{gensymb}

\newcommand{\unit}[1]{\,\mathrm{#1}}

\usepackage{amsmath}
\usepackage[euler]{textgreek}
\usepackage{graphicx}

\begin{document}

\preprint{APS/123-QED}

\title{Pump depletion and hot electron generation in long density scale length plasma with shock ignition high intensity laser}

\author{J. Li}
\affiliation{ 
Center for Energy Research, University of California San Diego, La Jolla, California 92093, USA
}%
\author{S. Zhang}%
\affiliation{
Center for Energy Research, University of California San Diego, La Jolla, California 92093, USA
}%

\author{C. M. Krauland}
\affiliation{%
Inertial Fusion Technology, General Atomics, San Diego, California 92121, USA
}%

\author{H. Wen}
\affiliation{ 
Department of Electrical Engineering, University of California, Los Angeles, California 90095, USA
}%

\author{F. N. Beg}%
\affiliation{ 
Center for Energy Research, University of California San Diego, La Jolla, California 92093, USA
}%

\author{C. Ren}
\email{chuang.ren@rochester.edu}
\affiliation{%
Department of Mechanical Engineering and Laboratory for Laser Energetics, University of Rochester, Rochester, New York 14627, USA
}%
\affiliation{Department of Physics and Astronomy, University of Rochester, Rochester, New York 14627, USA}

\author{M. S. Wei}
\altaffiliation[Current address:~]{Laboratory for Laser Energetics, University of Rochester, Rochester, New York 14627, USA}
\email{mingsheng.wei@rochester.edu}

\affiliation{%
Inertial Fusion Technology, General Atomics, San Diego, California 92121, USA
}%




\date{\today}

\begin{abstract}

Two-dimension Particle-in-cell simulations \textnormal{for laser plasma interaction with laser intensity of $10^{16} \unit{W/cm^2}$, plasma density range of 0.01-0.28$n_c$ and
scale length of $230 -330 \unit{\micro m}$} showed significant pump depletion of the laser energy due to stimulated
Raman scattering (SRS) and stimulated Brillouin
scattering (SBS) in the low density region ($n_e=0.01-0.2 n_c$). The simulations identified hot electrons
generated by SRS in the low density region with moderate energy and by two-plasmon-decay (TPD) near
$n_e=0.25n_c$ with higher energy. The overall hot electron temperature \textnormal{(46 keV)} and conversion efficiency \textnormal{(3\%)} were
consistent with the experiment measurements. The simulations also showed artificially reducing SBS would
lead to stronger SRS and a softer hot electron spectrum.

\end{abstract}

\pacs{Valid PACS appear here}
\maketitle



Shock ignition\cite{Betti2007ShockDensity,Perkins2009a} (SI) is a laser-driven inertial confinement fusion (ICF) scheme which achieves ignition conditions by using a low intensity ($<10^{15} \unit{W/cm^2}$) laser pulse for fuel compression followed by a high intensity ($10^{\sim16} \unit{W/cm^2}$) pulse.  This high intensity pulse drives a strong convergent shock in the dense shell to boost pressure and temperature to achieve ignition.
It potentially has much higher energy gain and lower risk of hydro instabilities\cite{Betti2007ShockDensity} compared to the conventional center hot spot scheme\cite{Lindl2004}. 
The challenge of SI lies in coupling sufficient ignition pulse energy to the target under 
significant laser plasma instabilities (LPI)\cite{KruerBook2003}, including stimulated Brillouin scattering (SBS)\cite{Liu1974,Forslund1975}, stimulated Raman scattering (SRS)\cite{Rosenbluth1972,Drake1973}, two-plasmon decay (TPD)\cite{RosenbluthM.N.WhitteR.B.Liu1973,Simon1983} and \textnormal{filamentation\cite{Craxton1984}}.
The LPI can scatter the laser \textnormal{light} to reduce the coupling efficiency (SBS and SRS), and 
generate suprathermal (hot) electrons (SRS and TPD) which can either preheat the fuel or enhance 
the coupling efficiency depending on their energy distribution\cite{LlorAisa2017}. Those hot electrons with energy below 100 keV can significantly boost the shock pressure\cite{Nora2015}. Recently Shang et al. showed for laser-hot electron conversion efficiency of $\eta>10\%$, ignition can be achieved with 400 kJ compression and 100 kJ ignition pulse energy in what they call {\it electron shock ignition} \cite{Shang2017}. 
Measuring and understanding LPI and hot electron generation in SI is critical\cite{Klimo2011,Weber2012,Batani2014,Riconda2011,Yan2014,Zhang2018,Hao2016,Hao2017,Li2017DensityTheory}. 

Existing experiment and simulation results on hot electron generation are somewhat conflicting. The strong spherical shock experiments on OMEGA measured $\eta$ up to 4\% \textnormal{with overlapped laser beams smoothed by spectral dispersion (SSD)} and 9\% without SSD \cite{Theobald2015} for the target density scale length at the quarter-critical surface $L_{n}=n_e/(dn_e/dx)=125\unit{\micro m}$\cite{Theobald2017}, where $n_e$ is the plasma density along the incident laser direction $x$. 
\textnormal{The instantaneous $\eta$} can be as high as 13-15\% \cite{Theobald2015,Theobald2017}. PIC simulations with the experimental conditions and the laser incident from \textnormal{plasma density} $n_e=0.12n_c$ yielded $\eta=12\%$\cite{Theobald2017}, where $n_c$ is the critical density of the incident laser. This motivated ignition-scale electron shock ignition design in \cite{Shang2017} with the expectation that $\eta$ would increase in longer scale lengths and stronger \textnormal{SRS}. The PIC simulations in \cite{Shang2017} with $L_n=314\unit{\micro m}$ and laser incident from $n_e=0.2n_c$ did show an $\eta=25\%$. On the other hand, the so-called 40+20 experiments on OMEGA using 40 beams for compression and 20 tightly-focused \textnormal{non-overlapping} beams as the ignition pulse measured an $\eta=1.7\%$ \cite{Theobald2012,Trela2018}. PIC simulations for this experiment with $L_n=170\unit{\micro m}$ and laser incident at $n_e=0.17n_c$ showed $\eta$ as high as 19\%, a clear discrepancy with the experiment. The PIC simulations did not include lower density region, which potentially can cause significant SRS and SBS backscttering. Indeed, 1D PIC \cite{Hao2016} and fluid \cite{Hao2017} simulations showed significant backscattering (of 40-90\%, depending on intensity, of the incident laser energy) once the $n_e=0.015-0.17n_c$ region was included. {\it It is critical to understand the low-density region LPI's since this region is long in ignition-scale targets.} 

\textnormal{Recently, a new SI experiment on OMEGA \textnormal{EP} using planar targets to achieve long-scale lengths \textnormal{($L_n=230-330\unit{\micro m}$)} measured $\eta=2\pm1\%$ with a hot electron temperature $T_h\sim 45$ keV  \cite{Zhang2018}. The back scattered SRS light spectra and the 4$\omega$ probe diagnostic measuring laser front movement indicated strong pump depletion in the region of $n_e<0.2n_c$ \cite{arXivZhang}. In this letter, we present 2-D, fully relativistic PIC simulations with physical conditions relevant to this experiment. To our knowledge, this was the first \textnormal{planar, multi-ps} 2-D PIC simulations of \textnormal{LPI at SI intensity} in \textnormal{millimeter long scale} plasmas with $n_e=0.01-0.28n_c$.
Our simulation results show strong pump depletion dominated by SBS, and $\eta=3\%$ and $T_h=46$ keV that were in excellent agreement with the experimental results. The observed hot electron properties were profoundly linked to SBS in the low density region, which can affect the location and saturation level of the most significant SRS modes through pump depletion. This new physics is unique to long scale-length plasmas and SI high intensity laser, for which the SBS seeds in the low density region can be amplified to the pump level and cause significant energy loss. Our results raise the concern of LPI-induced pump depletion in ignition-scale design for SI.} 

\begin{table}
\caption{\label{tab:table1} The density ranges and laser types for three PIC simulations.}
\begin{ruledtabular}
\begin{tabular}{lcr}
 Index & Density range($n_c$) & Laser type \\
  \hline
 i (Long)  & 0.01$\sim$0.28  & plane wave  \\
ii (Short)  & 0.14$\sim$0.28  & plane wave  \\
iii (Speckle)  & 0.14$\sim$0.28  & speckle  \\
\end{tabular}
\end{ruledtabular}
\end{table}

Our 2D PIC simulations, with the OSIRIS code\cite{Fonseca2002}, are listed in  Table I. The initial physical conditions \textnormal{of the simulations} were relevant to the OMEGA EP experiment \cite{Zhang2018} and obtained from hydro simulations with FLASH code\cite{Fryxell2000}. As shown in Fig.~1(a), the plasma density $n_e$ for Simulation i (Long), the main focus of this Letter, ranged from $0.01n_c$ to $0.28n_c$ with $L_n\sim230\unit{\micro m}$ for $n_e>0.2n_c$, and $L_n\sim330\unit{\micro m}$ for $n_e<0.2n_c$. \textnormal{This density profile had been validated in the experiment.} The initial plasma fluid velocity gradient was $1.5\times10^{-7} \unit{\omega_0}$, where $\omega_0$ is the laser frequency. The incident plane-wave ultraviolet(UV) laser (wavelength = 0.351$\unit{\micro m}$) was linearly polarized in the simulation plane with intensity $I_0=1\times10^{16}\unit{W/cm^2}$. It was launched at the left boundary of the simulation domain with short rising time of $10\omega_0^{-1}$ ($\sim1.9\unit{fs}$). The plasma consisted of electrons with temperature $T_e=1.6$ keV, Carbon and Hydrogen (CH) ions with temperature $T_i=1$ keV. The simulation was performed for 10 ps to demonstrate the time evolution and \textnormal{competition} of different LPI modes. The simulation domain was $16000 \unit{c/\omega_0}$($\sim900\unit{\micro m}$) (in longitudinal, $x$) by $200\unit{c/\omega_0}$($\sim11\unit{\micro m})$ (in transverse, $y$) with the grid sizes $\Delta x$ = $\Delta y$ = 0.2$\unit{c/\omega_0}$ ($\sim0.01\unit{\micro m}$), where $c$ is the light speed. The initial numbers of particles per cell were 100 for electrons and 50 for each ion species. The boundary conditions were periodic in $y$ direction. In $x$, the boundary conditions were thermal for particles and open for electromagnetic fields.

\begin{figure}
\includegraphics[width=3.375in]{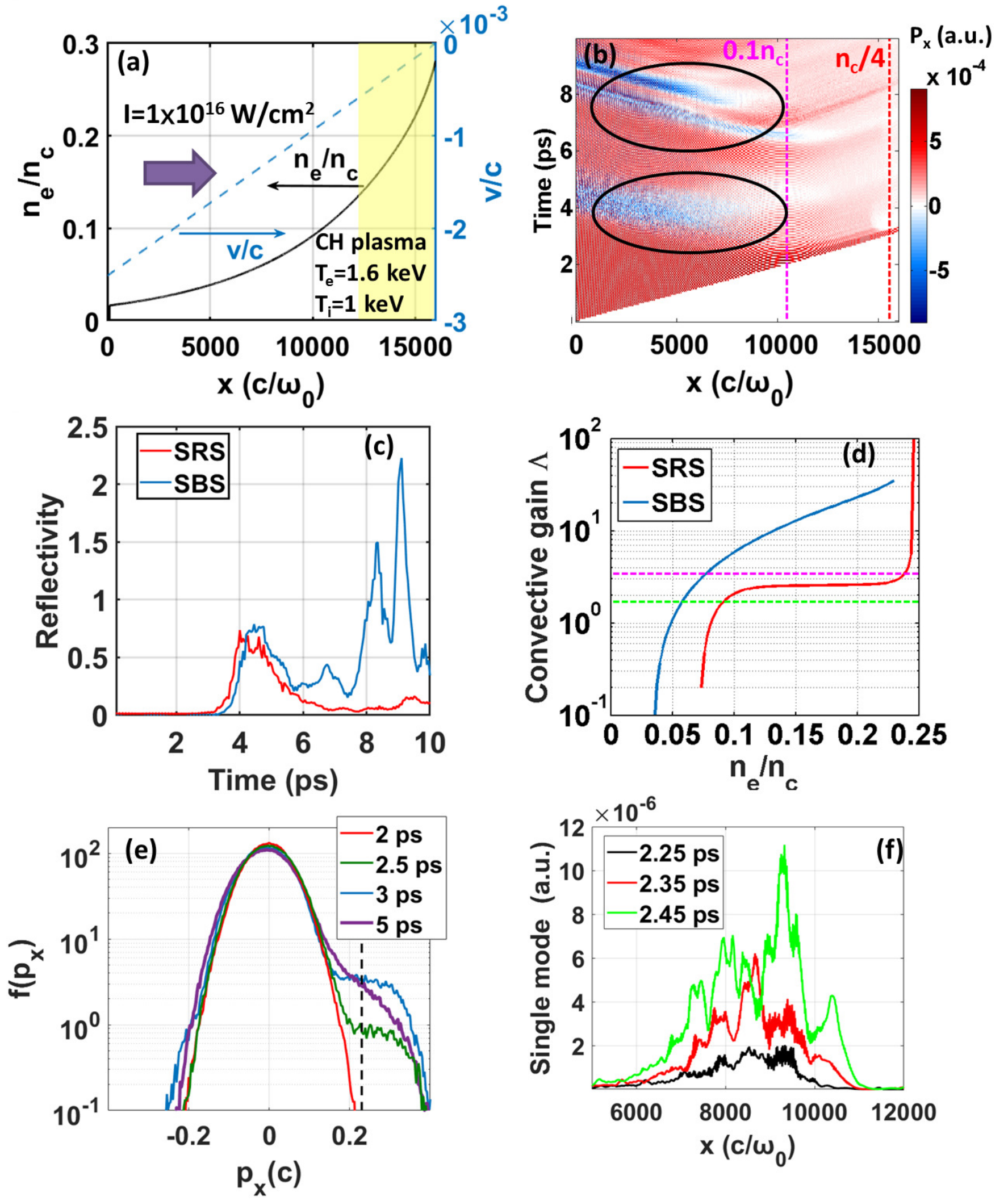}
\caption{\label{fig:ne_ve_pyt}(a) The initial density and fluid velocity profiles of the PIC simulations. (b) The space and time evolution of the longitudinal poynting vector $P_x$. (c) The instantaneous reflectivities of SRS and SBS. (d) The calculated $\Lambda_{\textnormal{SBS}}$ and $\Lambda_{\textnormal{SRS}}$ at different densities.\textnormal{The purple and green dashed lines refer to the $\Lambda_{\textnormal{pd}}$ for the Thomson scattering seed and PIC seed, respectively.} (e) The distribution functions of electron momentum near $0.1n_c$ at different times. The vertical dashed line marks the phase velocity of the resonant plasma wave of SRS modes. (f) The space profiles of the reflected SRS light with frequency $0.651\omega_0$.
}
\end{figure}

Significant pump depletion of the incident laser was observed from the $y$-averaged Poynting vector $P_x$-plot [Fig.~1(b)]. Strong negative $P_x$ bursts (blue color in the two black ellipses) correspond to significant back scattered light from SRS and SBS below 0.1$n_c$. The components of the bursts can be distinguished by the longitudinal wave number $k_x$ of the reflected lights near the left boundary with $n_e=0.01n_c$. Here, the SRS reflected lights had $0.5k_0<k_x<0.8k_0$, where $k_0$ is the incident laser wave number in vacuum. By comparison, the reflected lights from SBS had $k_x\sim k_0$. The SRS and SBS reflectivities are plotted in Fig.~1(c). At $\sim 4 \unit{ps}$, the instantaneous SRS reflectivity ($\sim75\%$) was comparable to SBS in the first burst, but SBS ($66\%$ \textnormal{on average}) strongly suppressed SRS ($8\%$ \textnormal{on average}) after 6 ps. 

The strong laser pump depletion was due to the large convective gains of SRS and \textnormal {especially, SBS. In an inhomogeneous plasma, the linear convective gain\cite{RosenbluthM.N.WhitteR.B.Liu1973}  $\Lambda=\gamma_{\textnormal{eff}}^2/|\kappa\prime V_1V_2|$ describes how much SRS and SBS can grow from an initial \textnormal{$\delta$-function} seed. 
Here $\gamma_{\textnormal{eff}}$ is approximated by $\sqrt{(\gamma_0-\nu_1)(\gamma_0-\nu_2)}$, where $\gamma_0$ is the temporal growth rates of SRS or SBS in a homogeneous plasma without any damping, $V_1$, $V_2$, $\nu_1$ and $\nu_2$ are the group velocities and total damping rates on the two daughter waves. We consider both Landau damping and collisional damping\cite{KruerBook2003}. The Landau damping only applies to the plasma wave for SRS and ion acoustic wave for SBS\cite{Vu1994AnPlasma}, and the collisional damping affects both daughter waves for SRS and SBS. The gradient of the wave number mismatch is $\kappa\prime=d(k_{\textnormal{pump}}-k_1-k_2)/dx$, where $k_{\textnormal{pump}}$, $k_1$ and $k_2$ are the wave numbers of the pump and two daughter waves. Large $I_0$ ($I_0\propto\gamma_0^2$) and $L_n$ ($L_n\propto1/\kappa\prime$) lead to large $\Lambda$. Since the scattered light intensity $I_{\textnormal{scatter}}\approx \exp{(2\pi\Lambda})I_{\textnormal{seed}}$ ($I_{\textnormal{seed}}$ is the seed intensity), one can reasonably estimate that strong pump depletion occurs when  $I_{\textnormal{scatter}}$ becomes comparable to the incident laser intensity $I_0$. For the actual $I_{\textnormal{seed}}\approx(1-8)\times10^{-10}I_0$ (based on the Thomson scattering model \cite{Hao2016,Berger1989}for SBS), the critical gain above which pump depletion by SBS becomes important was $\Lambda_{pd}^{(SBS)}=(2\pi)^{-1}ln(I_0/I_{seed})= 3.4$. (The SRS seed level was lower, $I_{\textnormal{seed}}\approx(5-6)\times10^{-11}I_0$.) } 


\textnormal{We plot $\Lambda$ for SRS and SBS in Fig.~1(d). The results show that SBS gain factor $\Lambda_{\textnormal{SBS}}$ was above $\Lambda_{pd}^{(SBS)}$ for $n>0.06n_c$ while the gain factor for SRS $\Lambda_{\textnormal{SRS}}$ stayed at $\sim2$. This suggests that SBS, not SRS, can cause significant pump depletion \textnormal{for $n_e<0.1n_c$}. Both gains significantly decreased at lower densities due to strong Landau damping.}   \textnormal{We did observe flattening of the electron velocity distribution [Fig.~1(e)] which can cause decreasing of Landau damping. Neglecting all the damping effects, $\Lambda_{\textnormal{SRS}}=2.7$  which was still much lower than $\Lambda_{\textnormal{SBS}}$.} This was consistent with the observed suppression of SRS caused by SBS in the region near $0.1n_c$ [Fig.~1(b)(c)].

However, previous theory predicted SRS may turn from convective to absolute at shock ignition intensities \cite{Li2017DensityTheory} by density modulations \cite{Nicholson1974,Nicholson1976,Williams1979,Picard1985DecayInstability}, potentially growing SRS beyond the convective limit. According to \cite{Li2017DensityTheory} the density modulation threshold for this transition is low, $\Delta n/n\sim10^{-5}$ for a characteristic modulation length of $\sqrt{V_1V_2}/\gamma_0=25 c/\omega_0$. This is well below the typical 2\% in our simulations and also measured in experiments with similar parameters  \cite{Moody1999}.  Here we present evidence that SRS may have turned absolute in our PIC simulations. At $0.1n_c$, the resonant mode of the SRS back scattered light had a frequency of $\omega_{\textnormal{SRS}}=0.651\omega_0$. The wave numbers of this mode can be calculated at different densities, enabling the extraction of this mode amplitude in space. Fig.~1(f) show the space profiles of this mode at 2.25, 2.35 and 2.45 ps near $0.1n_c$ ($x=10500c/\omega_0$). The chosen time frames were right after the laser reached $0.1n_c$ at 2ps but before strong pump depletion due to SBS reduces the laser intensity by half at 2.5 ps. The left peaks near $x=8000c/\omega_0$ displayed signatures of convective growth, growing in amplitude and moving to the left.  In contrast, the right peak in the resonant region near $x=10000 c/\omega_0$ kept growing at the same location, showing the signatures of absolute instability. Its growth eventually was interrupted by SBS-induced pump depletion that lowers the local pump intensity. This illustrates the importance of LPI coupling in this region. 



Pump depletion due to SBS and SRS reduced the laser intensity near $n_c/4$ to $\sim8\times10^{14}\unit{W/cm^2}$, only $\sim8\%I_0$. But this intensity was still well above the threshold for absolute TPD ($1.6\times10^{14}\unit{W/cm^2}$)\cite{Simon1983}. TPD signals are presented in Fig.~2. 
Near $n_c/4$, significant plasma waves was observed [Fig.~2(a)], and the Fourier transform of these modes showed that they overlapped with the theoretical curve of TPD instability[Fig.~2(b)] above 0.2$n_c$. \textnormal{Note that the strong signals near $k_x\sim0.9\unit{\omega_0/c}$ in Fig.~2(b) corresponded to the electric fields of both the incident laser and the SBS side scattered light.}   

\begin{figure}
\includegraphics[width=3.375in]{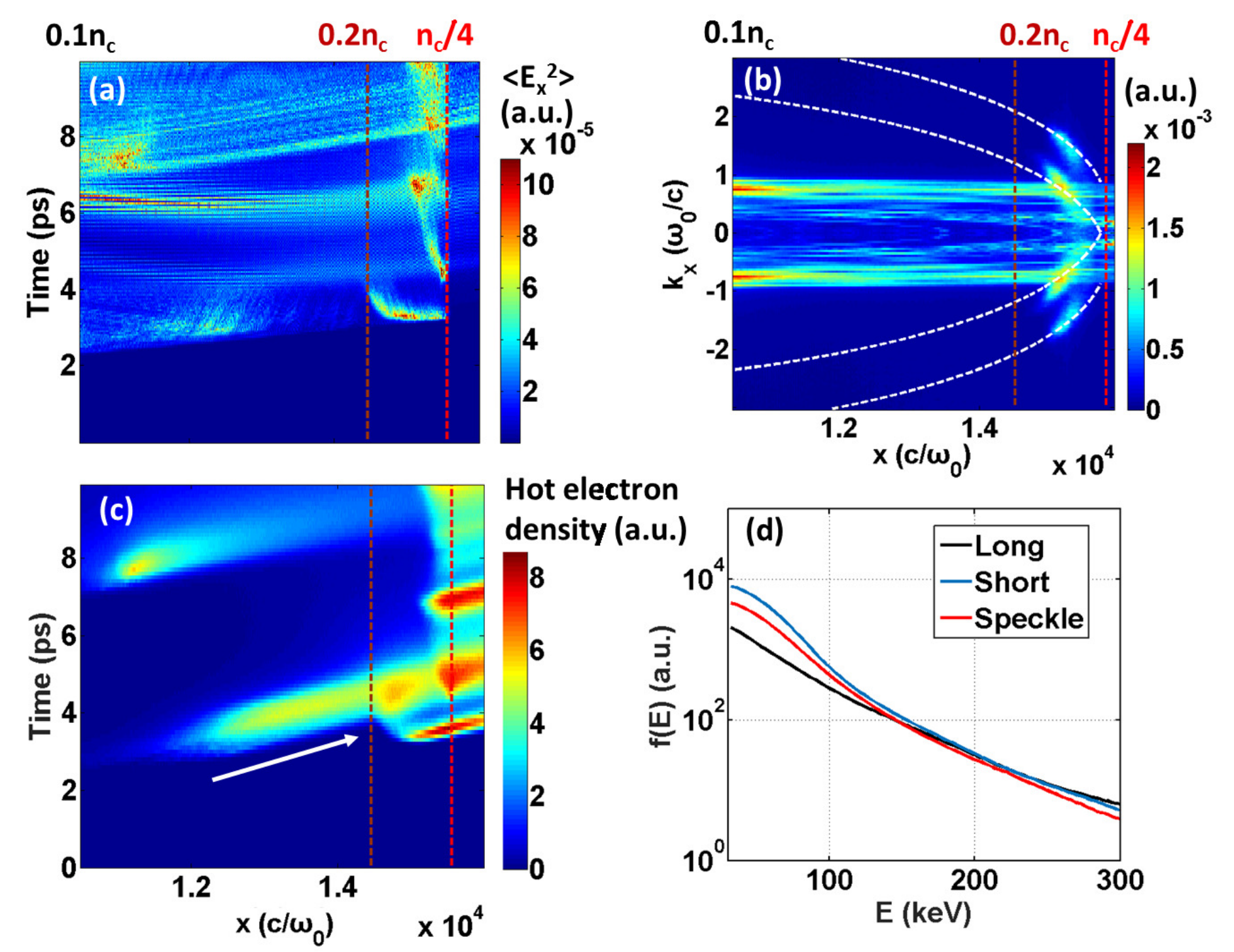}
\caption{\label{fig:hot_electron} (a) The space and time evolution of longitudinal electric field energy $<E_x^2>$. (b) The $k_x-x$ spectrum of $E_x$ at 10 ps. The white curves represent the $k_x$ corresponding to the maximum TPD growth rate at different densities. (c) The density of hot electrons with energy above 50 keV in the phase space of $x$ and time. The white arrow shows the hot electrons move to higher density region with time. (d) \textnormal{The energy spectra of accumulated hot electrons from Simulation (i)-(iii) (respectively denoted by ``Long", ``Short" and ``Speckle" in the legend).} In (a), (b) and (c), the left boundaries correspond to the plasma density 0.1$n_c$.
 }
\end{figure}

The space-time information of hot electrons (with energy above 50 keV) is presented in Fig.~2(c) and shows that most hot electrons were generated in $n_e=0.2-0.25n_c$ and TPD may be the main cause. However, in $n_e=0.1-0.2n_c$, SRS-generated hot electrons were also observed at 4 ps and 8 ps. The SRS hot electrons may continue to be accelerated by TPD [Fig.~2(c)]. To our knowledge this was the first time both the SRS and TPD hot electrons were identified by their origins in one simulation. The time-averaged $\eta$ was $3\%$ with a temperature of $T_h\sim 46 \unit{keV}$[Fig.2~(d) (Long)]. 

\textnormal{The high SBS reflectivity was not due to elevated seed level in the simulation. In the simulation, the SBS seed was dominated by the electromagnetic noise, which was about $10^3$ times higher than electrostatic noise.   The PIC SBS seed level was higher than the actual seed level $I_{\textnormal{seed}}^{(PIC)}\approx10^{-4\sim-5}I_0 \approx10^{5}I_{seed}$. However, the critical gain $\Lambda_{pd}^{(SBS)}$ depends {\it insensitively} on the seed level, and would decrease from 3.4 to 1.7. This did not expand significantly the density region of strong SBS pump depletion [Fig. 1(d)]. This was further supported by two 1D PIC simulations with the same physical parameters but different number of particles per cell PPC=200 and 20,000, which showed similar average SBS reflectivity [Fig. 3(a)]. We believe here physics depended more on strong pump depletion and not sensitive to the seed level\cite{Tang1966}.}

\textnormal { The experiment found strong pump depletion but did not have direct SBS measurement \cite{Zhang2018}.} 
To study possible situations when SBS was not as high \textnormal{below $0.1n_c$} and a higher laser intensity can reach $n_e>0.1n_c$, we performed two more 2D PIC Simulation (ii \& iii) [Table I] with the same initial conditions except that the density profile starts from 0.14$n_c$ instead of 0.01$n_c$ [the yellow region of Fig.~1(a)]. \textnormal{Both simulation showed a lower $T_{h}$ than the experiment.} Without the low density ($<0.14n_c$) plasma, the average SRS reflectivity for the plane wave case [ii, (Short)] rose to 20$\%$ and the SBS reflectivity decreased to 32$\%$ as shown in Fig.~3(b). Hot electrons were now mainly generated by SRS in $n_e=0.14-0.2n_c$ [Fig.~3(c)], which contrasts to Fig.~2(c). TPD can still be observed near $n_c/4$ but they were no longer the main cause for hot electrons. Compared to Simulation (i),  $\eta$ increased from $3\%$ to $5\%$ [Fig.~3(d)] and $T_h$ decreased from 46 keV to 35 keV [Fig.~2(d)], both trending away from the experiment values. 

\textnormal{Comparing Simulation (i) and (ii) shows the importance of including the low density region. This also shows the competition in SBS and SRS. Reducing SBS not only reduced the overall reflectivity but also increased SRS, which in turn would soften the hot electron spectrum. From hot electron locations in the simulation domain [Fig.~ 2(c) and 3(c)], we can separate SRS and TPD hot electrons. We find SRS hot electrons had $T_h=20\sim$25 keV, lower than $T_h=45\sim65$ keV of TPD hot electrons. Consequently, the accumulated hot electrons in Simulation (i) showed higher $T_h\sim$ 46 keV compared to moderate $T_h\sim$ 35 keV in Simulation (ii) [Fig.~2(d)].} \textnormal{This shows the critical effects of LPI in the low density region on the hot electron properties in the long scale-length plasma.}

\begin{figure}
\includegraphics[width=3.375in]{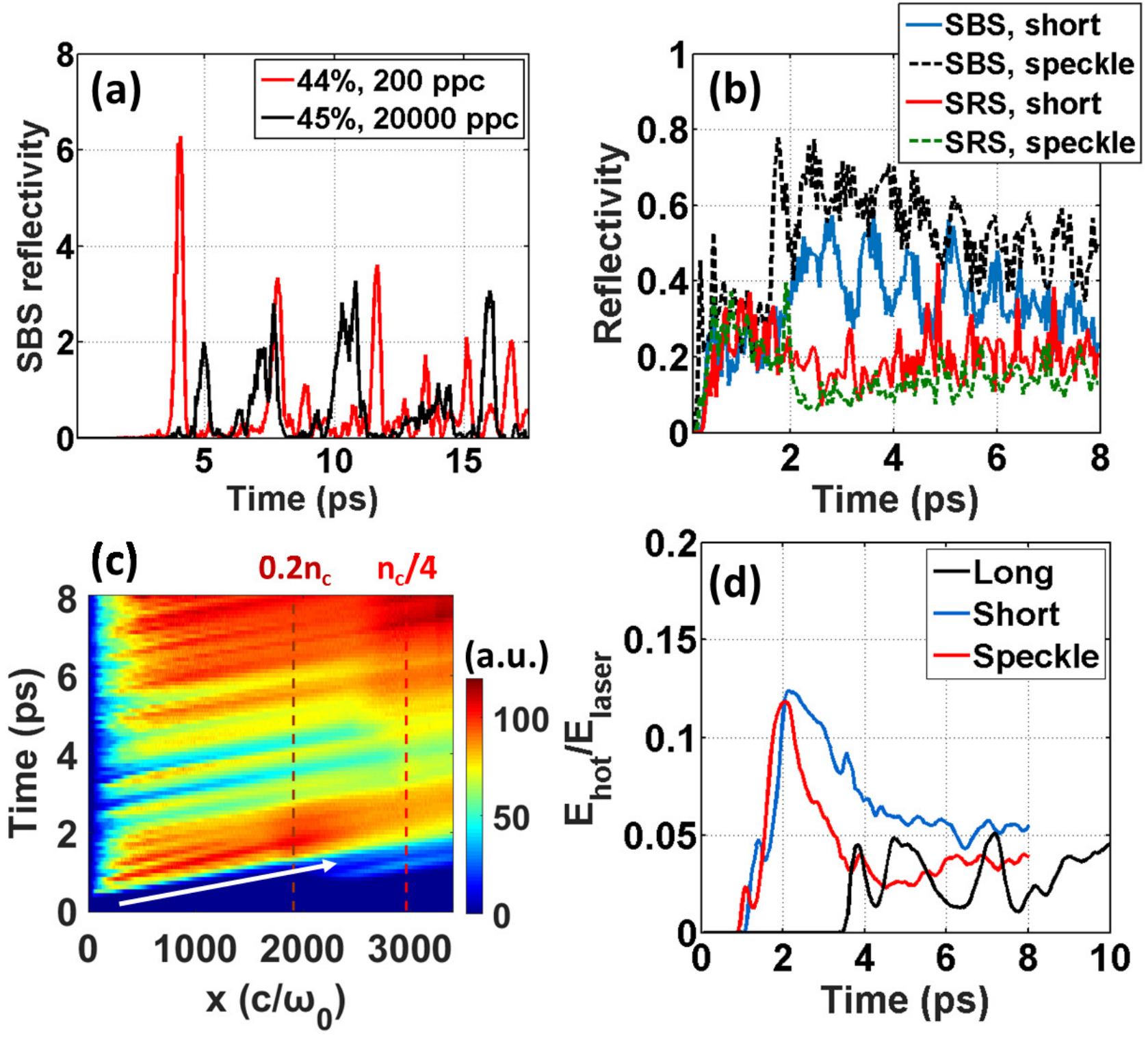}
\caption{\label{fig:small} \textnormal{(a) The SBS reflectivities in 1D PIC simulations with numbers of particles per cell (ppc) = 200 and 20000. The time-averaged reflectivities are given in the legend.} (b) The reflectivities of SRS and SBS in Simulation (ii) and (iii) denoted by ``Short" and ``Speckle" . (c) The density of hot electrons with energy above 50 keV in the phase space of $x$ and time. (d) The instantaneous $\eta$ for Simulation (i), (ii) and (iii).  
 }
\end{figure}


\textnormal{We performed Simulation (iii) [Table I] to qualitatively study how the effects of potential laser filaments/speckles on SBS and hot electron generation.} We used a single Gaussian laser speckle focused at $n_c/4$ with a spot size of 3$\unit{\micro m}$, and the transverse size of the simulation domain was doubled compared to Simulation (ii). The maximum laser intensity at the incident plane was $I\sim2\times10^{16}\unit{W/cm^2}$,  keeping the transversely averaged intensity at $1\times10^{16}\unit{W/cm^2}$. All other initial conditions were the same with the Simulation (ii). The SRS and SBS reflectivities are plotted in Fig.~3(b). Compared to the plane wave case (ii), the SBS reflectivity increased from $32\%$ to 50$\%$ and SRS reflectivity decreased from $20\%$ to 15$\%$. The conversion efficiency reduced to $3\sim4\%$ [Fig.~3(d)] but $T_h=37$ keV did not change much [Fig.~2(d)].


In summary, the large-scale PIC simulations here show the importance of the low density region in shock ignition LPI hot electron generation. Only when it was included can the simulations reproduce the experiment measurements of the hot electron conversion efficiency and temperature. Excluding it would increase the conversion efficiency by reducing the overall pump depletion and lower the temperature by increasing the SRS hot electron fraction to levels inconsistent to the experiments. \textnormal{The strong pump depletion was supported by the experiments} \cite{arXivZhang}. Our research shows that high convective gains of LPI for SI high laser intensity in the long-scale low density region is a concern. However, it should be noted that both the simulations and experiments described in this paper used a single-beam as the main interaction pulse. Recent planar target experiments on OMEGA performed by this team\cite{Zhang2018} using multiple and overlapped UV beams as the interaction pulses with $I_0=1\times10^{16} W/cm^2$, $L_n\sim230\micro m$ and $T_e=3 keV$ have doubled laser-to-electron energy conversion efficiency $\eta=4\pm2\%$ compared to the single-beam experiment on OMEGA EP. This warrants further investigation. Future LPI experiments on OMEGA are planned to directly probe electron plasma and ion acoustic waves using optical Thomson scattering together with time-resolved full aperture backscattering diagnostics. 

\textnormal{
We acknowledge useful discussions with Dr.~L. Hao. This work was supported by the DOE Office of Science under grant No.DE-SC0014666, DE-SC0012316 and NNSA NLUF grant No.DE-NA0003600, \textnormal{DE-NA0002730}. \textnormal{This research used resources of the National Energy Research Scientific Computing Center (NERSC) and the Argonne Leadership Computing Facility, which are both the U.S. Department of Energy Office of Science User Facility, but operated respectively under Contract No.DE-AC02-05CH11231 and DE-AC02-06CH11357}. The support of DOE does not constitute an endorsement by DOE of the views expressed in this paper.}

\bibliography{main.bbl}

\end{document}